\documentclass[12pt]{article}
\usepackage{amssymb}
\begin{document}
\newtheorem{prop}{}[section]
\newtheorem{defi}[prop]{}
\newtheorem{lemma}[prop]{}
\newtheorem{rema}[prop]{}
\def\parn{\par \noindent}
\def\fine{\hfill $\diamond$ \vskip 0.2cm \noindent}
\newcommand{\rref}[1]{(\ref{#1})}
\def\beq{\begin{equation}}
\def\feq{\end{equation}}
\def\parn{\par\noindent}
\def\lt{$LT$ }
\def\vain{\rightarrow}
\def\interi{{\textbf Z}}
\def\lag{Lagrange }
\def\kow{Kowalewski }
\def\bh{bH }
\def\qbh{qbH }
\hyphenation{di-men-sio-nal}
\newcommand{\CH}{{\cal H}}
\newcommand{\CP}{{\cal P}}
\newcommand{\CD}{{\cal D}}
\newcommand{\CA}{{\cal A}}
\newcommand{\CW}{{\cal W}}
\newcommand{\CZ}{{\cal Z}}
\newcommand{\CS}{{\cal S}}
\newcommand{\CU}{{\cal U}}
\newcommand{\CN}{{\cal N}}
\newcommand{\CV}{{\cal V}}
\newcommand{\CM}{{\cal M}}
\newcommand{\CQ}{{\cal Q}}
\newcommand{\CL}{{\cal L}}
\newcommand{\CK}{{\cal K}}
\newcommand{\del}{{\partial}}
\def\lag{Lagrange }
\def\kow{Kowalewski }
\def\bh{bH }
\def\qbh{qbH }

\makeatletter
\@addtoreset{equation}{section}
\renewcommand{\theequation}{\thesection.\arabic{equation}}
\makeatother
\begin{titlepage}
\begin{center}
{\huge The quasi-bi-Hamiltonian formulation of the Lagrange top}
\end{center}
\vspace{1truecm}
\begin{center}
{\large
Carlo Morosi${}^1$,~Giorgio Tondo${}^2$  } \\
\vspace{0.5truecm}
${}^1$ Dipartimento di Matematica, Politecnico di
Milano, \\ P.za L. da Vinci 32, I-20133 Milano, Italy \\
e--mail: carmor@mate.polimi.it \\
${}^2$ Dipartimento di Scienze Matematiche, Universit\`a di Trieste, \\
via A. Valerio 12/1, I-34127 Trieste, Italy \\
e--mail: tondo@univ.trieste.it \\
\end{center}
\vspace{1truecm}
\begin{abstract}
\noindent
Starting from the tri-Hamiltonian formulation of the Lagrange top in
a six-dimensional phase space, we
discuss the possible reductions  of the
Poisson tensors, the vector field and its Hamiltonian functions on a
four-dimensional  space. We
show that the vector field of the Lagrange top possesses, on the
reduced phase space, a quasi-bi-Hamiltonian formulation, which
provides a set of separation
variables  for the corresponding Hamilton-Jacobi equation.
\end{abstract}
\vspace{0.4truecm} \noindent
\end{titlepage}
\section{Introduction}
\label{intro}
The classical theory of separation of variables for the
Hamilton-Jacobi equation provides
the most effective tool to solve the equations of motion of a given
Hamiltonian system. In this framework, the
main problem is to have an efficient (possibly
algorithmic) way to {\it
produce} a set of separation variables.  To this purpose, two new approaches,
stemming from soliton theory, have been  recently introduced: the
``magic Sklyanin's
recipe"\cite{Sk},  based on the Lax representation  of the equations
of the motion, and the
bi-Hamiltonian ($bH$) approach to separation of variables \cite{bro, mt,
bla, FMPZ}, based on
the bi-Hamiltonian structures associated with the equations of motion.
A remarkable feature of the latter approach is that if the
Hamiltonian system admits
a {\em{quasi-bi-Hamiltonian}} ($qbH$) formulation,
then a set of separation
variables can be algorithmically computed \cite{mt}; moreover, the
$qbH$ property is independent
of the coordinate system  in which the $bH$
structure is written down.
\par\noindent
The aim of this paper is to apply the approach based on the $qbH$
property to the classical
Lagrange top $(LT)$; in particular, we show how  the (complex) separation
variables for $LT$, introduced in \cite{med} in an algebraic-geometric setting,
arise quite naturally as distinguished functions for its
tri-Hamiltonian structure.
\par\noindent
The starting point of our analysis is the fact that, on a
six-dimensional phase
space $M$, the \lt vector field $X_L$ admits a tri-Hamiltonian formulation
$X_L= P_\alpha dh_\alpha$ (throughout the paper, the index $\alpha$
takes values
$0, 1, 2$), each one of the three compatible Poisson tensors
$P_\alpha$ possessing two
independent Casimir functions.
\parn
When one tries to eliminate the Casimirs by fixing their values, one
is faced with
a typical situation, occurring also for other $bH$
finite-dimensional
integrable systems \cite{GT1, FMT, FMPZ}: to each one of the
symplectic leaves $S_\alpha$
one can restrict only the vector field  $X_L$ and the corresponding pair
($P_\alpha, h_\alpha$), but not the entire triple of the Poisson
tensors, so that the
tri-Hamiltonian formulation of $X_L$ is lost under restriction.
Nevertheless, using a more general reduction process {\em \`a la
Marsden-Ratiu}, we will
show that the symplectic leaf $S_0$ of the Poisson tensor $P_0$ can
be endowed with a
Poisson-Nijhenuis structure~\cite{MM,FoFu} (hence a $bH$ structure) and
that $X_L$ can be given a $qbH$ formulation.  So, the separability of
$LT$ is obtained from
its Hamiltonian structures as a natural outcome of the reduction process.
\parn
The paper is organised as follows. In Section \ref{rev} the
tri-Hamiltonian structure of \lt is shortly reviewed; in Section
\ref{reminder} the main
properties of the $qbH$ model are discussed in view of application to $LT$. In Sections
\ref{reduno} and \ref{reddue}, respectively, the reduction of the
Poisson tensors $P_{\alpha}$ and of the vector field $X_L$ with its Hamiltonian
functions are considered; the $qbH$ formulation for $X_L$ is explicitly
constructed, together with a solution of the corresponding
Hamilton-Jacobi equation. Our results are summarised in Section 6, where some
potential extensions of this work are pointed out.
\vskip 0.4cm
\noindent
\section{The multi-Hamiltonian structure of the Lagrange top}
\label{rev}
A modern formulation of $LT$ can be found in
\cite{aud, gav}; as  usual in this framework, the components of vectors and
covectors and the entries
of matrices are referred to the comoving frame, whose axes are the
principal inertia
axes of the top, with fixed point $O$. \parn
The phase space $M$ of \lt is parametrised by the pair $m=(\omega,\gamma)$,
where $\omega = (\omega_1, \omega_2, \omega_3)^T~$  and $\gamma =
(\gamma_1, \gamma_2,
\gamma_3)^T~$ are the angular velocity and the vertical unit vector,
respectively. The following
notations are introduced: $\mu$ is the mass of the top, $g$ the
acceleration of gravity, $J=\mbox{diag}(A, A, cA)$ the principal inertia matrix
$(c\neq 1)$, $G=(0,0,a)^T$ is the center of mass; at last,
normalisations are chosen so
that $\mu a g/A=1$.
\par\noindent
  The Euler-Poisson equations are $dL_o/dt=M_o$ (change of the angular momentum)
and $d\gamma/dt=0$ (invariance of the vertical unit vector); with the above
notations and normalisations, these equations take the well-known form
\beq
{dm\over dt}=X_L(m)~,\qquad
X_L(m)= \left(
\begin{array}{c}
(1-c) \omega_2 \omega_3-\gamma_2 \\
-(1-c) \omega_3 \omega_1+\gamma_1 \\
0\\
\gamma_2 \omega_3- \gamma_3 \omega_2\\
\gamma_3 \omega_1- \gamma_1 \omega_3\\
\gamma_1 \omega_2- \gamma_2 \omega_1\\
\end{array}
\right)~.
\label{moto}
\feq
The \lt vector field $X_L$ can be given a tri-Hamiltonian formulation
\beq
X_L= P_0 dh_0= P_1 dh_1= P_2 dh_2~;
\label{tri}
\feq
the compatible Poisson tensors $P_\alpha$, written in matrix
block form, are
\beq
{P}_0= \left(
\begin{array}{cc}
0  & B \\
B & C \\
\end{array}
\right)~,
\qquad
{P}_1= \left(
\begin{array}{cc}
-B  & 0 \\
0 & \Gamma \\
\end{array}
\right)~,
\qquad{P}_2= \left(
\begin{array}{cc}
T  & R \\
-R^T & 0 \\
\end{array}
\right)~,
\label{poisson}
\feq
where $B$, $C$, $\Gamma$, $T$ and $R$ are $3\times 3$ matrices
$$
B= \left(
\begin{array}{ccc}
       0  & -1 & 0 \\
1 & 0 & 0 \\
0 & 0& 0 \\
\end{array}
\right),
\quad
C= \left(
\begin{array}{ccc}
       0  & c ~\omega_3 & -\omega_2 \\
- c ~\omega_3 & 0 & \omega_1 \\
\omega_2 & -\omega_1 & 0 \\
\end{array}
\right),
\quad
\Gamma = \left(
\begin{array}{ccc}
       0  & \gamma_3 & -\gamma_2 \\
-\gamma_3 & 0 & \gamma_1 \\
\gamma_2 & -\gamma_1 & 0 \\
\end{array}
\right),
$$
\beq
T= \left(
\begin{array}{ccc}
       0  & - c~ \omega_3 & \omega_2/ c \\
       c~ \omega_3 & 0 & - \omega_1 / c \\
- \omega_2 / c & \omega_1/ c & 0 \\
\end{array}
\right)~,
\quad
R = \left(
\begin{array}{ccc}
       0  & -\gamma_3 & \gamma_2 \\
\gamma_3 & 0 & -\gamma_1 \\
-\gamma_2/ c & \gamma_1 /c & 0 \\
\end{array}
\right) ~.
\label{bcgamma}
\feq
The Hamiltonian functions $h_\alpha$ can be written as
\beq
h_0={1\over 2} F_4+2 \sigma c F_1 F_3,\qquad
h_1=\sigma c^2  F_{1}^3- F_3-2 \sigma c F_1 F_2,\qquad h_2=F_2,
\label{ham}
\feq
where $\sigma= {c-1\over 2c }$ and
\beq
F_1=\omega_3,\qquad\qquad  F_2={1\over 2} (\omega_{1}^2+\omega_{2}^2+
c~ \omega_{3}^2)
-\gamma_3,
\label{integrali}
\feq
$$
F_3= \omega_1\gamma_1+\omega_2\gamma_2+ c~ \omega_3\gamma_3~,\qquad
F_4= \gamma_{1}^2+\gamma_{2}^2+ \gamma_{3}^2~.
$$
As it is known, the functions $F_i$ $(i=1,...,4)$ are integrals of motion for
Eq.\rref{moto}; they are independent and in involution w.r.t. each
one of the three Poisson tensors. Moreover, $(F_1, F_2)$ are Casimir
functions of $P_0$,
$(F_1, F_4)$ of $P_1$ and $(F_3, F_4)$ of $P_2$. \parn
The vector field $X_L$ can be immersed in two different $bH$ chains,
starting and ending with the Casimirs of the Poisson tensors $P_\alpha$:
\beq
P_0 dF_2=0,\quad P_2 dF_2=P_0 dh_0=X_L, \quad P_2 dh_0=P_0
d(-\sigma F_3^2),
\quad P_2 d(-\sigma F_3^2)=0~;
\feq
$$
P_0 dF_2=0,\quad P_1 dF_2=P_0 dh_1, \quad P_1 dh_1=P_0 dh_0=X_L, \quad
P_1 dh_0= P_0 d(-\sigma c F_1 F_4),
$$
$$
\quad P_1 d( -\sigma c F_1 F_4)=0~.
$$
\vskip 0.2cm \noindent
\begin{prop}
\label{rem1}
{\bf{Remark.}}  The Hamiltonian formulation of \lt w.r.t. $P_2$ is
classical (see, e.g.,\cite{gav}).
The \bh formulation w.r.t. $(P_0, P_2)$ was introduced in \cite{rat}
in  the semidirect product $\mathfrak{so}(3)\times \mathfrak{so}(3)$, and
was later recovered in   \cite{med} in an algebraic-geometric  setting.
The tri-Hamiltonian formulation w.r.t. $(P_0, P_1, P_2)$ was
constructed in \cite{mag},
by a suitable reduction of the Lie-Poisson pencil defined in the
direct sum of three copies of
$\mathfrak{so}(3)$. (To compare the above-quoted results, let us
recall that the angular
momentum and the vertical unit vector are taken as dynamical
variables in \cite{gav,rat,mag},
whereas the angular momentum is replaced by the angular velocity
$\omega$ in \cite{med} and in the
present paper.)
\fine
\end{prop}
\vskip 0.4cm
\noindent
\section{The quasi-bi-Hamiltonian model}
\label{reminder}
The $qbH$ model was introduced in \cite{cab,bro}
and developed in \cite{mt,tm} (see also \cite{bla} and
references therein). Here we summarise some facts to be used in the
rest of the paper.\parn
Let $Q_0$, $Q_1$ be two compatible Poisson tensors on a manifold $M$;
a vector field $X$ is said to admit a $qbH$ formulation w.r.t. $Q_0$
and $Q_1$ if there
are three functions $\rho$, $H$, $K$ such that
\beq
X= {Q}_0 ~dH= {1\over \rho} ~{Q}_1 ~dK~.
\label{ro}
\feq
In other words, $X$ is Hamiltonian w.r.t. $Q_0$ with Hamiltonian
function $H$, and it is {\em quasi-Hamiltonian}
($qH$) w.r.t. $Q_1$, with $qH$ function $K$ and
conformal factor $1/\rho$. In spite of the presence
of $\rho$, equation (\ref{ro}) implies that $H$ and $K$ are in involution
w.r.t. both Poisson brackets corresponding to $Q_0$ and $Q_1$~(as well
as in the particular $bH$ case $\rho=1$).
\parn
If $\mbox{dim}~M=2n$, the $qbH$ formulation is said to be of maximal
rank if at each point $m\in M$ the Poisson tensors $Q_0$, $Q_1$ are
non degenerate and the associated tensor $N=Q_1~ Q_{0}^{-1}$
(with vanishing Nijenhuis torsion) has $n$ independent
eigenvalues $\lambda_1(m),..., \lambda_n(m)$. In this case, one
can introduce a local chart
$(\lambda_i, \mu_i)$~$(i=1,2,...,n)$, called  a Darboux-Nijenhuis
chart \cite{mm}, such that $Q_0$,  $Q_1$ and $N$ take the canonical form
\beq
Q_0=~\left(
\begin{array}{cc}
0 & I_n \\
-I_n& 0 \\
\end{array}
\right),
\quad
Q_1=~\left(
\begin{array}{cc}
0 & \Lambda \\
-\Lambda & 0 \\
\end{array}
\right),\qquad
\quad
N=~\left(
\begin{array}{cc}
\Lambda &0 \\
0&\Lambda  \\
\end{array}
\right)~,
\label{diag}
\feq
with $\Lambda=\mbox{diag} (\lambda_1,...,\lambda_n)$; in general, the
coordinate functions $\mu_i$, canonically conjugated to
$\lambda_i$, can be computed by quadratures. At last, the $qbH$
formulation is said to be of Pfaffian type if $\rho=\prod_{i=1}^n \lambda_i$.
\parn
The following result has been proved in \cite{mt} for a Pfaffian $qbH$
vector field.
\begin{prop}
\label{hkgen}
\textbf{Proposition.} The general solution of Eq.\rref{ro} for the
Pfaffian case is given by functions $H$ and $K$ which, in a
Darboux-Nijenhuis chart
$(\lambda_i, \mu_i)$, take  the ``canonical'' form
\beq
H=\sum_{i=1}^n {f_i \over \Delta_i}, \qquad
K=\sum_{i=1}^n {\rho\over \lambda_i}~{f_i \over \Delta_i}~,
\qquad\quad
\Delta_i= \prod_{j\neq i} (\lambda_i-\lambda_j)~,
\label{gen}
\feq
where each $f_i$ is an  arbitrary  function, depending at most
on the pair $(\lambda_i, \mu_i)$.
Moreover, the Hamilton-Jacobi equations for both $H$ and $K$ are separable.
\fine
\end{prop}
\vskip 0.2cm \noindent
This Proposition has a straightforward consequence.
\begin{prop}
\label{rem2}
{\textbf{Corollary.}} Let $X=Q_0~dH$ be a Hamiltonian vector field;
if in a $Q_0$-Darboux
chart $(x,y)$ the Hamiltonian $H$ takes the canonical form \rref{gen},
then $X$ admits  a Pfaffian $qbH$ formulation w.r.t. a Poisson tensor $Q_1$
and a $qH$ function $K$ of the form \rref{diag} and \rref{gen},
respectively.
\parn
Viceversa, let $X=1 / \rho~ Q_1~dK$ be a $qH$ vector field w.r.t.  $Q_1$;
if, in a chart $(x,y)$, $Q_1$ and $K$ take the canonical forms
\rref{diag} \rref{gen}
and $\rho=\prod_{i=1}^n x_i$,
then it is also $X=Q_0~dH$ with $Q_0$ and $H$ given by (\ref{diag})
(\ref{gen}), respectively. Hence,  the chart
$(x,y)$ is a Darboux-Nijenhuis chart for the Poisson pair $Q_0, Q_1$.
\fine
\end{prop}
For $n=2$, this  Corollary can be slightly generalised, in a way that
is useful for
subsequent applications to \lt.
\begin{prop}
\label{hkgen1}
\textbf{Proposition.} Let $S$ be a four-dimensional manifold and
$Y=Q_0~ dH $ be a
Hamiltonian vector field w.r.t. a non degenerate Poisson tensor
$Q_0$. Let there is a Darboux chart $(x, y)$ such that the
Hamiltonian $H$ can be written
as a linear combination of two functions $\hat{H}$, $\hat{K}$ with
the canonical form
\rref{gen}, i.e.,
\beq
H(x,y)= \beta \hat{H}(x,y) + \hat{K}(x,y) \qquad \beta=const ~, \label{y}
\feq
$$\hat{H}(x,y)={1\over x_1-x_2} \left(\hat{f}_1(x_1, y_1)-
\hat{f}_2(x_2, y_2)\right)~,
$$
$$
\hat{K}(x,y)={1\over x_1-x_2} \left(x_2 \hat{f}_1(x_1, y_1)- x_1
\hat{f}_2(x_2, y_2)\right)~.
$$
Then, the vector field $Y$ admits the Pfaffian $qbH$ formulation
(\ref{ro})-(\ref{gen});
a Darboux-Nijenhuis chart $(\lambda,\mu)$ is given by the following map:
\beq
\Phi:(x,y)\mapsto (\lambda,\mu)\qquad\quad
\lambda_i={1\over  x_i+\beta}~,\quad \mu_i=-y_i (  x_i+\beta)^2
\qquad(i=1,2)~.
\label{fi}
\feq
Hence, $H$ is separable in the chart  $(\lambda,\mu)$.
Moreover, $H$ is separable also in the chart $(x,y)$  and the corresponding
Hamilton-Jacobi equation
\beq
H(x_1, x_2, \partial W /\partial x_1, \partial W /\partial x_2)=h
\label{hamjac}
\feq
has the complete solution $ W(x_1, x_2; \hat{h},\hat{k})= W_1(x_1;
\hat{h},\hat{k})+
W_2(x_2; \hat{h},\hat{k})$~, $W_1$ and $W_2$ fulfilling the Sklyanin
separation equations
\cite{Sk}
\beq
\hat{f}_1(x_1, W'_1(x_1))= x_1 \hat{h} -\hat{k}, \qquad
\hat{f}_2(x_2, W'_2(x_2))= x_2 \hat{h} -\hat{k} \ ,
\label{eqSk}
\feq
with $\beta\,\hat{h} + \,\hat{k}= h~$.
\end{prop}
{\textbf{Proof.}}
It is straightforward to check that the map $ \Phi: (x,y)\mapsto
(\lambda,\mu)$ is
a Darboux map for $Q_0$; moreover, since
$x_1-x_2=-(\lambda_1-\lambda_2) /{ \lambda_1 \lambda_2}$,
the Hamiltonian $H$ takes the canonical form (\ref{gen}):
\beq
H \Big(x(\lambda,\mu),y(\lambda,\mu)\Big)=
\beta \hat{H} \Big(x(\lambda,\mu),y(\lambda,\mu)\Big) +
\hat{K} \Big(x(\lambda,\mu),y(\lambda,\mu)\Big)=
\feq
$$= { 1\over \lambda_1-\lambda_2} \left( - \lambda_1
\hat{f}_1  ({1\over\lambda_1} -\beta,
-\lambda_1^2\mu_1 )  +
\lambda_2 \hat{f}_2  ({1\over\lambda_2} -\beta,
-\lambda_2^2 \mu_2) \right)=
$$
$$
= {1\over \lambda_1-\lambda_2} \Big(f_1(\lambda_1, \mu_1)
-f_2(\lambda_2, \mu_2)\Big)~,
$$
where
\beq
f_1(\lambda_1 , \mu_1)=-\lambda_1\hat{f}_1 ({1\over\lambda_1} -\beta,
-\lambda_1^2\mu_1)~,
\quad
f_2(\lambda_2 , \mu_2)=-\lambda_2\hat{f}_2  ({1\over\lambda_2} -\beta,
-\lambda_2^2\mu_2)~.
\feq
On account of Corollary \ref{rem2}, the vector field
$Y= Q_0 dH$ admits the $qH$ formulation $Y=1/\rho~ Q_1 dK$ and $H$ is
separable.
\parn
Obviously enough, $H$ is separable also in the chart $(x,y)$, since
the map $\Phi$ is a {\it separated}  map \cite{Be}, i.e., it maps
separated coordinates
into separated ones.
Indeed, taking into account the form \rref{y} of the function $H$, it
is easily checked that the Hamilton-Jacobi equation $H(x,\partial
W/\partial x)= h$  has a
complete solution
$ W(x_1, x_2; \hat{h}, \hat{k})=$ $W_1(x_1; \hat{h}, \hat{k})+
W_2(x_2; \hat{h}, \hat{k})$, with $\beta\,\hat{h} +\,\hat{k}= h~$, and that
$W_1$, $W_2$ fulfil the Sklyanin separation equations
(\ref{eqSk}) for the Hamilton-Jacobi equations $\hat{H}(x,\partial
W/\partial x)=\hat{h}$, ~
$\hat{K}(x,\partial W/\partial x)=\hat{k}$.
\fine
\section{The reduction of the tri-Hamiltonian structure of the Lagrange top}
\label{reduno}
If a vector field $X$ on a manifold $M$ is  $bH$ w.r.t. a
pair of degenerate Poisson tensors $(P_0,  P_1)$,
a preliminary step in analysing its integrability is  trying to reduce
the vector field,
its Hamiltonian functions and the Poisson tensors on a
lower-dimensional manifold $M'$, where one of
the two Poisson tensors, say $P_0$, be invertible. A natural way to
do that is to fix the values of
the Casimir functions of $P_0$. Of course, both $P_0$ and $X$ can be
properly {\em  restricted} to
a symplectic leaf $S_0$, giving rise to a Poisson tensor $P'_0$ and
to a vector field
$X'= P'_0~dH'$, $H'$ being the  restriction to $S_0$ of the
original Hamiltonian $H$.
However, without additional assumptions, $P_1$ is not assured to
restrict to $S_0$, so that $X'$ loses the original $bH$ formulation.
\parn
This situation occurs also for the tri-Hamiltonian structure of \lt.
Each one of the three
Poisson tensors $P_\alpha$ has two independent Casimir functions, and
the generic
symplectic leaves $S_\alpha$ are four-dimensional submanifolds of $M$.
On account of Eq.\rref{integrali}, they are defined as
\beq
S_0=\{ m\in M~\vert~ \omega_3={a_1 \over 2c},~~
\omega_{1}^2+\omega_{2}^2+ c \omega_{3}^2-2\gamma_3=2a_2 \}~,
\label{leave0}
\feq
$$
S_1=\{ m\in M~\vert~ \omega_3={a_1 \over 2c},
~~\gamma_{1}^2+\gamma_{2}^2+  \gamma_{3}^2=a_4\}~,
$$
$$
S_2=\{ m\in M~\vert~  \omega_1\gamma_1+\omega_2\gamma_2+ c
\omega_3\gamma_3= -{1\over
2}a_3~,~~\gamma_{1}^2+\gamma_{2}^2+  \gamma_{3}^2=a_4\}~,
$$
where $a_1$, $a_2$, $a_3$ and $a_4$ are arbitrary  constants.
Each Poisson tensor $P_\alpha$ can be properly restricted to a corresponding
symplectic leaf $S_\alpha$, but the other two tensors do not restrict
to the same leaf.
\parn
Nevertheless, a quite general reduction technique given by the Marsden-Ratiu
theorem ~\cite{MR} can be applied; it will enables us to construct
on $S_\alpha$
a Poisson-Nijhenuis structure~\cite{MM,FoFu} induced by the
tri-Hamiltonian structure on $M$, and on $S_0$ a $qbH$ formulation for
the vector field $X_L'$.
Essentially, one considers
a Poisson manifold $(M, P)$, a submanifold $S\hookrightarrow M$
and a distribution $D\subset TM_{\vert_S}$ such that
$E:=D\cap TS$ is a regular foliation with a good quotient
$\CN= S/E$.
Then, the theorem states that the Poisson tensor $P$ is reducible to $\CN$ if
the following conditions hold:
\parn
{\bf i)} the functions on $M$ which are invariant along $D$ form a
Poisson subalgebra
of $C^\infty(M)$;
\parn
{\bf ii)} $P(D^\circ)\subset TS + D$ ($D^\circ$ being the annihilator of $D$
in $T^*M$).
\parn
Analogously to previous applications of this procedure to $bH$
structures \cite{FMT, FMPZ, FMP},
let us choose as the submanifold $S$ a generic symplectic leaf
$S_\alpha$ of the Poisson tensor $P_\alpha$ and a distribution $D$
such that at each
point $s_\alpha \in S_\alpha$ the following decomposition holds:
\begin{equation} \label{DTS}
{T_{s_{\alpha}}} M={T_{s_{\alpha}}} S_\beta \oplus {D_{s_{\alpha}}} \ ,
\end{equation}
$S_\beta$ being the symplectic leaf of $P_\beta \ (\beta=0,1,2)$ passing
through $s_\alpha $.
\parn
This assumption assures that {\bf ii)} is trivially fulfilled and that $E=0$,
so that the reduction procedure becomes a {\em submersion} $\Pi: M
\rightarrow S_\alpha$
onto the manifold $S_\alpha$;  then,  it allows us to endow
$S_\alpha$ with a  non degenerate tri-Hamiltonian structure, since the kernels
of the reduced Poisson tensors $P'_\beta  $ vanish.
Indeed, if $\Pi^*$ denotes the (injective) pull-back of the
submersion $\Pi$, it is
\beq
{Ker_{s_{\alpha}}} P'_\beta=(\Pi^*)^{-1} \Big({Im_{s_{\alpha}}}\Pi^*\cap
P_\beta^{-1}({D_{s_{\alpha}}}\cap{T_{s_{\alpha}}} S_\beta)\Big)
\feq
$$\stackrel{
(\ref{DTS})}{=}
(\Pi^*)^{-1}({Im_{s_{\alpha}}}\Pi^*\cap {Ker_{s_{\alpha}}} P_\beta)=0  ,
$$
where we have taken into account that
\beq
{Im_{s_{\alpha}}}\Pi^*\subset D^\circ, \qquad
D^\circ \cap {Ker_{s_{\alpha}}} P_\beta=D^\circ \cap
{(Im_{s_{\alpha}}} P_\beta)^\circ =D^\circ \cap
{(T_{s_{\alpha}}}S_\beta )^\circ\stackrel{ (\ref{DTS})}{=}0.
\feq
In the $LT$ case, the distribution is as follows.
\begin{prop}\label{ld:lem}
\textbf{Lemma.}
Let $D$ be the distribution given by the vector fields
\begin{equation}
Z_1=-i c\frac{\del}{\del \omega_2}+
\frac{\del}{\del \omega_3},  \qquad
Z_2=i\frac{\del}{\del \gamma_2}-
\frac{\del}{\del \gamma_3}\
\end{equation}
($i=\sqrt{-1}$). Moreover, let $\varphi_1, \varphi_2$ be two
generic functions. Then, for each Poisson tensor $P_\alpha$ there are
two vector fields
$W_{1\alpha}$ and $W_{2\alpha}$
(depending on $\varphi_1$ and $\varphi_2$) such that
\begin{equation} \label{eq:lieZP}
L_{\varphi_1 Z_1+ \varphi_2Z_2}\,(P_\alpha)=Z_1\wedge
W_{1\alpha}+Z_{2}\wedge W_{2\alpha}
\end{equation}
($L_Z$ and $\wedge$ denoting the Lie derivative along the flow of the vector
field $Z$ and the exterior product of vector fields, respectively).
\end{prop}
{\bf Proof.}
It is easy to check that $L_{ Z_j}P_\alpha= Z_1\wedge Y_{1j\alpha}+Z_{2}\wedge
Y_{2j\alpha}$ ($j=1,2$), with suitable vector fields $Y_{1j\alpha},
Y_{2j\alpha}$.
This result, together with the identity $L_{f X}(P)=fL_{X}(P)+X\wedge
P\,df$, implies
(\ref{eq:lieZP}), the vector fields $W_{j\alpha}$ being
$W_{j\alpha}=\varphi_1 Y_{j1\alpha} +\varphi_2
Y_{j2\alpha}+P_\alpha d\varphi_j$.
\fine
\parn
Eq.(\ref{eq:lieZP}) implies the assumption {\bf i)}, since
if $f$ and $g$
are invariant functions along $D$ and $Z \in D$, then $L_{\varphi Z}\{f,g\}
=<df, L_{\varphi Z}(P) dg>\stackrel {(\ref{eq:lieZP})}{=}0$ for each function
$ \varphi$. Moreover, condition (\ref{DTS})  is generically satisfied as
it can be easily verified.  Hence,  conditions
{\bf i)}, {\bf ii}) are fulfilled and the Marsden-Ratiu reduction
technique can be applied on each symplectic leaf $S_\alpha$. In
conclusion, we have proved the following.
\begin{prop}
\label{prpo4}
\textbf{Proposition.}
The tri-Hamiltonian structure $P_\beta$ is {\em reducible} to a non degenerate
tri-Hamiltonian structure $P'_\beta$ on each one of the symplectic
leaves $S_\alpha$.
\fine
\end{prop}
To express the reduced tensors in a particularly simple
and useful form, it is convenient to adapt the coordinates on $M$ to
the distribution $D$, introducing a parametrisation including
coordinate functions which span
the subalgebra of the functions invariant along $D$.
Let us choose the chart $(u, v, w)$, related to $(\omega, \gamma)$ by the map
$\Psi:M \rightarrow M:$ $(\omega, \gamma)\mapsto (u, v, w)$
\beq
\label{map}
u_1=c\omega_3-i \omega_2~,\quad u_2=i\gamma_2-\gamma_3~,
\feq
$$
v_1=\omega_1,\quad v_2=-\gamma_1,\qquad w_1=i\omega_2+c \omega_3,\quad
w_2=-i\gamma_2-\gamma_3~.
$$
Taking into account the tri-Hamiltonian structure $P_\alpha$ given by
\rref{poisson} and the definition \rref{leave0} of $S_\alpha$, a
straightforward (though lengthy) calculation allows one to verify that
the chart ($u, v$) gives a parametrisation on each one
of the symplectic leaves $S_\alpha$; the reduced Poisson
tensors $P'_{\beta}$ and the tensor $N$ take the form
\begin{equation}
\label{eq:P0}
P'_0=i\left(
\begin{array} {cccc}
0& 0& 0&1  \\
0& 0& 1&  u_1  \\
0& -1& 0& 0\\
-1& - u_1&0& 0
\end{array}
\right) \ ,
\qquad
P'_1=i\left(
\begin{array} {cccc}
0& 0& 1& 0 \\
0& 0&0&-u_2\\
-1 & 0& 0& 0\\
0 &u_2& 0&0
\end{array}
\right) \ ~,
\end{equation}
$$
P'_2=i\left(
\begin{array} {cccc}
0& 0& -u_1& -u_2 \\
0& 0& -u_2& 0 \\
u_1 & u_2& 0& 0\\
u_2& 0& 0&0
\end{array}
\right) \ .
$$
\begin{prop}
\textbf {Remark.}
\label{hankel}
By a direct inspection, one easily concludes that the tensor
$N':=P'_1 {P'_0}^{-1}$
(with vanishing Nijenhuis torsion) is such that
$P'_1= N' P'_0$ and $P'_2=N' P'_1$.
\par\noindent
The matrix representation of $P'_0$ and  of the adjoint tensor ${N'}^*$
of $N'$ are formed by Hankel and Frobenius blocks, respectively, so
that $(u,v)$ are
Hankel-Frobenius coordinates, in the terminology of \cite{FMT}.
\fine
\end{prop}
\begin{prop}
\label{conclude}
\textbf{Proposition.}
Let us consider the map $\Psi: S_\alpha \rightarrow S_\alpha:$ $(u,
v)\mapsto (x, y) $
\beq
\label{another}
x_1={1\over 2} (-u_1-\sqrt{u_{1}^2-4u_2})~,\qquad
x_2={1\over 2} (-u_1+\sqrt{u_{1}^2-4u_2})~,
\feq
$$
y_1={1\over 2} (2 v_2-u_1 v_1- v_1 \sqrt{u_{1}^2-4u_2})~,\qquad
y_2={1\over 2} (2 v_2-u_1 v_1+v_1 \sqrt{u_{1}^2-4u_2})~.
$$
The chart $(x,y)$ is a Darboux-Nijenhuis chart for the
tri-Hamiltonian structure on
$S_\alpha$; the reduced Poisson tensors $P'_\alpha$ have the
matrix block form
\beq
P'_0=-i~\left(
\begin{array}{cc}
    0 & I \\
-I& 0 \\
\end{array}
\right)~\qquad
P'_1=-i~\left(
\begin{array}{cc}
0 & \cal X \\
-\cal X & 0 \\
\end{array}
\right)~\qquad
P'_2=-i~\left(
\begin{array}{cc}
0 & {\cal X} ^2 \\
-{\cal X }^2& 0 \\
\end{array}
\right)~,
\label{ppp}
\feq
where $ {\cal X} =\mbox{diag} (x_1, x_2)$~.
\end{prop}
{\bf{Proof.}} A straightforward computation, taking into account
Eq.s \rref{eq:P0} and \rref{another}.
\fine
(To be more precise, in
order to have the Darboux-Nijenhuis chart defined in Section
\ref{reminder} one should
eliminate the factor $(-i)$ in Eq.\rref{ppp}, via the map $x \mapsto i
x$, $y \mapsto y)$.
\vskip 0.4cm
\section{The reduction of the vector field and the Hamiltonians of
the Lagrange top.}
\label{reddue}
Having established the projection of the tri-Hamiltonian structure on
each one of the
symplectic leaves $S_\alpha$, the next step is to consider the
reduction of the vector
field $X_L$ and of the corresponding Hamiltonian functions $h_\alpha$.
\parn
Unfortunately, they  {\it do not project} onto $S_\alpha$, since
$X_L$ does not preserve the
distribution $D$ and the Hamiltonians $h_\alpha$  are not invariant
along $D$; hence, the tri-Hamiltonian formulation of $X_L$ is lost on
$S_\alpha$.
Nevertheless, each pair $(X_L, h_\alpha)$ can be {\it restricted} to
the corresponding
symplectic leaf $S_\alpha$, so that Eq.\rref{moto}, restricted to
$S_\alpha$, keeps a
Hamiltonian formulation. Furthermore, if we consider  the reduction
on a symplectic
leaf $S_0$, we can recover, as a reminder of the original tri-Hamiltonian
formulation, a $qbH$ formulation for $X_L$; this suffices to provide a
set of separation
variables. Indeed, the following holds.
\begin{prop}
\label{trivial}
\textbf{Proposition.} The vector field $X_L$, restricted to $S_0$,
takes the form
\beq
X_L=P'_0 dH=-i~Q_0~ dH~.
\label{cinqueuno}
\feq
Its Hamiltonian $H={h_0}_{\vert S_0}$ takes the form
\beq
H(x, y)= \sigma a_1~ \hat{H}(x, y)+ \hat{K}(x, y) ~,
\label{cinquedue}
\feq
where
\beq
\hat{H}(x, y)={1\over x_1-x_2}\left(\hat{f}(x_1, y_1)-\hat{f}(x_2,
y_2)\right)~,
\label{cinquetre}
\feq
$$
\hat{K}(x, y)={1\over x_1-x_2}\left(x_2 \hat{f}(x_1, y_1)-x_1
\hat{f}(x_2, y_2)\right)~,
$$
$$
\hat{f}(\xi, \eta)= -{1\over 2}~ \eta^2 + {1\over 2}~ \xi^4 + {1\over
2}~ a_1 \xi^3 +
(a_2+\sigma~{a_{1}^2\over 4})\xi^2~.
$$
\end{prop}
{\bf{Proof.}} A straightforward computation.
\fine
On account of this result, we are just in the situation considered in
Proposition \ref{hkgen1}, with
\beq
\beta=\sigma a_1, \qquad \hat{f}_1=\hat{f}_2=\hat{f}~.
\feq
So, ${X}_L$ admits a $qbH$ formulation; the Darboux-Nijenhuis
coordinates $(\lambda, \mu)$
are obtained
from $(x, y)$ via the map \rref{fi}:
\beq
\lambda_i=\left(x_i+ \sigma a_1\right)^{-1}, \qquad
\mu_i=- y_i~\left(x_i+ \sigma a_1\right)^2\qquad (i=1,2).
\label{trinv}
\feq
As it follows from the general results of Propositions \ref{hkgen},
\ref {hkgen1},
$H$ and $K$ are separable both in the Darboux-Nijenhuis chart
$(\lambda, \mu)$ and in the
chart $(x,y)$. Using the latter, let us compute
a  solution $W$ of the Hamilton-Jacobi equations for $H$ and $K$
\beq
H(x_1,x_2, {\partial W\over \partial x_1},
{\partial W\over \partial x_2})= h, \qquad
K(x_1, x_2,  {\partial W\over \partial x_1},
{\partial W\over \partial x_2})= k ;
\label{HK}
\feq
taking into account the expression \rref{cinquetre} of $\hat{f}$ and
the fact that the $qH$
function $K$ given by \rref{gen} turns out to be  $K=\hat{H}$, we have
\beq
W(x_1,x_2; h,k)= \int^{x_1} \sqrt{g(\xi)}~ d\xi+
\int^{x_2} \sqrt{g(\xi)}
~ d\xi~,\feq
$$
g(\xi)=  \xi^4+ a_1 \xi^3 + (2 a_2 +\sigma {a_{1}^4\over2} ) \xi^2
- 2 k \xi + 2(h-\sigma a_1 k).
\label{wx}
$$

\section{Concluding remarks}
The first result in this paper is that, reducing {\em \`a la Marsden-Ratiu}
the tri-Hamiltonian structure ($P_0$, $P_1$, $P_2$) of $LT$ onto a
generic symplectic
leaf $S_\alpha$ of each Poisson tensor, a non degenerate Poisson-Nijenhuis
structure  is obtained. The reduction depends
essentially on the distribution $D$ fulfilling (\ref{DTS}) and
(\ref{eq:lieZP}); since $D$ may be
not unique, possibly different Poisson-Nijenhuis structures  can be
constructed on the symplectic leaf. This point deserves further investigations.
\par\noindent
The second step of the reduction procedure is the restriction of the
$LT$ vector field
and Hamiltonian functions to
the invariant submanifold $S_0$, discussed in Section 5. This produces a
quasi-bi-Hamiltonian formulation for the $LT$ vector field  and
consequently, as a
necessary outcome, a set of separation variables. An open question
is whether the
restriction of the $LT$
vector field to other invariant submanifolds, such as the symplectic
leaves $S_1$ and
$S_2$ of the Poisson tensors $P_1$ and $P_2$, gives rise to different sets
of separation variables.
\par\noindent
As a last remark, we observe that the tri-Hamiltonian structure of $LT$ has a
deformation in the original phase space $M$ (see, e.g., \cite{mag}).
In fact, there is a
vector field $\tau$ such that
$L_\tau(P_2) = 2 P_1$, $L_\tau(P_1) = P_0$, $L_\tau(P_0) = 0; $
in the chart $(\omega, \gamma)$ chosen in this paper, $\tau$ is given by
$\tau=(0,0,-2/c, \omega_1,\omega_2,c\, \omega_3)^T$. On the contrary,  a
recursion operator $N$ relating the Poisson tensors does not exist in $M$.
Once the reduction onto the submanifold $S_0$ has been performed   under 
the submersion $\Pi: M \rightarrow S_0$, the deformation
process is preserved since the vector field $\tau$ is projectable onto $S_0$. Hence, 
the previous relations hold for
$P'_0,~P'_1,~P'_2$ w.r.t. the
projected vector field $\tau'$, given by $\tau'=-(2, u_1, 0, v_1,)^T$
in the chart $(u, v)$.
As  observed in Remark \ref{hankel}, on $S_0$ there
is also an
invertible recursion operator $N'$ such that $P'_1 = N' P'_0$ and
$P'_2 = N' P'_1$
(consequently, one has a
whole sequence of compatible Poisson tensors $P'_{j+1}=N' P'_j$ for
each integer $j~$).
One may wonder whether the recursion scheme based on $N'$ could be
inferred from the existence
of the deformation scheme on the initial phase space, and under which
conditions on the
deformation vector field $\tau$~. At the best of
our knowledge, this question (which is not peculiar of $LT$ only) has
not yet received a
satisfactory answer; in our opinion, it  deserves further investigations in
the general framework of the reduction theory for multi-Hamiltonian manifolds.
\vskip 1cm\noindent
{\bf Acknowledgments.}
This work was partially supported by Italian M.I.U.R., under the
research project {\it Geometry of
Integrable Systems}, and by INDAM (G.N.F.M.). We thank an anonymous
referee for useful suggestions about
the style of the paper.

\end{document}